\documentclass[12pt,preprint]{article}
\usepackage{amssymb}
\usepackage{amsmath}
\usepackage{graphics}
\usepackage{epsfig}

\newcommand{\eqb}{\begin{equation}}
\newcommand{\eqe}{\end{equation}}
\newcommand{\dmb}{\begin{displaymath}}
\newcommand{\dme}{\end{displaymath}}

\newcommand{\eab}{\begin{eqnarray}}
\newcommand{\eae}{\end{eqnarray}}

\newcommand{\be}{\begin{equation}}
\newcommand{\ee}{\end{equation}}

\begin{document}

\begin{titlepage}

\vspace{0.6cm}

\begin{center}
{\Large {Light scalars as tetraquarks: decays and mixing with quarkonia} 
\vspace{1.0cm} }

{\large {Francesco Giacosa\footnote{%
Talk given at the International School of Nuclear Physics, 29th Course, `$%
Quarks$ $in$ $Hadrons$ $and$ $Nuclei$', Erice-Sicily, 16-24 September 2007.
E-mail: giacosa@th.physik.uni-frankfurt.de}} }
\end{center}

\vspace{2.0cm}

\begin{center}
\emph{Institut f\"ur Theoretische Physik\\[0pt]
Universit\"at Frankfurt\\[0pt]
Johann Wolfgang Goethe - Universit\"at\\[0pt]
Max von Laue--Str. 1\\[0pt]
60438 Frankfurt, Germany\\[0pt]
}
\end{center}

\vspace{2.0cm}

\begin{abstract}
The tetraquark assignement for light scalar states below 1 GeV is discussed
on the light of strong decays. The next-to-leading order in the large-N
expansion for the strong decays is considered. Mixing with quarkonia states
above 1 GeV is investigated within a chiral approach and the inclusion of
finite-width effects is taken into account.
\end{abstract}

\end{titlepage}

\setlength{\baselineskip}{0.9\baselineskip}

\section{Introduction}

The nature of scalar states, under debate since over 30 years, represents
one of the major problems of low energy QCD, see \cite{revs} for reviews. In
these proceeding, based on the works \cite{tq,tqmix,lupo}, the tetraquark
hypothesis for the light scalar mesons is discussed.

The resonances $a_{0}(980),$ $f_{0}(980)$ and $f_{0}(600)$ are well
established \cite{pdg} and evidence for the state $k(800),$ although not yet
decisive, is mounting. Thus, a full nonet emerges below 1 GeV. Why not to
interpret it as a quarkonium scalar nonet? Some serious problems are well
known: (a) the mass degeneracy of $a_{0}^{0}(980)$ and $f_{0}(980)$, which
would be $\sqrt{1/2}(\overline{u}u-\overline{d}d)$ and $\overline{s}s$ in
the $\overline{q}q$ assignment, is unexplained. (b) The coupling of $%
a_{0}(980)$ to $\overline{K}K$ is large and points to a hidden s-quark
component. (c) Scalar quarkonia are p-wave (and spin 1) states, therefore
expected to lie above 1 GeV as the tensor and axial-vector mesons. (d)
Quenched lattice results \cite{quenched} find a quarkonium isospin 1 mass $%
M_{u\overline{d}}=1.4$-$1.5$ GeV, thus showing that $a_{0}(1450),$ rather
than $a_{0}(980)$, is the lowest scalar $I=1$ quarkonium; recent unquenched
results of Ref. \cite{graznew} which find in addition to $a_{0}(1450)$ also $%
a_{0}(980)$, are discussed in the conclusions. (e) In the large-$N_{c}$
study of Ref. \cite{pelaez} it is shown that the resonance $f_{0}(600)$ does 
$not$ behave as a quarkonium or a glueball state: the width does not scale
as $1/N_{c}$ and the mass is not constant.

These problems can be solved in the framework of the tetraquark assignment
of the light scalars: as shown by Jaffe 30 years ago \cite{jaffeorig}, when
composing scalar diquarks in the color and flavor antitriplet configuration
(good diquarks) instead of quarks, the resonance $f_{0}(980)$ is dominantly
\textquotedblleft $\overline{s}s(\overline{u}u+\overline{d}d)$%
\textquotedblright\ and the neutral isovector $a_{0}(980)$ as
\textquotedblleft $\overline{s}s(\overline{u}u-\overline{d}d)$%
\textquotedblright , thus neatly explaining the problem (a) mentioned above.
The $f_{0}(600)$ is the lightest state with dominant contribution of
\textquotedblleft $\overline{u}u\overline{d}d$\textquotedblright , in
between one has the kaonic state $k(800)$ ($k^{+}$ interpreted as
\textquotedblleft $\overline{d}d\overline{s}u$\textquotedblright ): the mass
pattern is nicely reproduced. This is still one of the most appealing
properties of the tetraquark assignment. Support for the existence of
Jaffe's states below 1 GeV is also in agreement with the Lattice studies of
Refs. \cite{jaffelatt}. Concerning the other problems mentioned above: (b)
is solved because $a_{0}(980)$ has a hidden s-quark content. Points (c) and
(d) are solved setting the quarkonia states above 1 GeV: $a_{0}(1450)$ and $%
K(1430)$ are the isovector and isodoublet respectively, $f_{0}(1370)$, $%
f_{0}(1500)$ and $f_{0}(1710)$ are the isoscalar states with glueball's
intrusion (with $f_{0}(1500)$ being a hot candidate) \cite{glue}. Point (e)
is solved in virtue of the large-$N_{c}$ counting for tetraquarks: widths
and masses increase for increasing $N_{c}$ \cite{jaffenew}.

In the following we concentrate on quantitative aspects of the tetraquark
assignment: in Section 2 the Clebsch-Gordan coefficients for the strong
decays are studied, in Section 3 the inclusion of mixing with quarkonia
states is investigated within a chiral approach; subsequently, the inclusion
of mesonic loops and finite-width effects is analyzed. Finally, in Section 4
the conclusions are presented.

\section{Strong decays in a $SU_{V}(3)$-invariant approach}

In the original work of Jaffe \cite{jaffeorig} the decay of tetraquark
states takes place via the so-called superallowed OZI-mechanism (Fig. 1.a):
the switch of a quark and an antiquark generates the fall-apart of the
tetraquark into two mesons. This mechanism is dominant in the large-$N_{c}$
counting. Already in \cite{jaffeorig} the possibility that a quark and an
antiquark annihilate is mentioned in relation to isoscalar mixing but is not
explicitly evaluated for the decay rates. The fact that only one
intermediate transverse gluon is present, see Fig. 1.b, may indicate that
this mechanism, although suppressed of a factor $1/N_{c}$, is relevant. In 
\cite{maiani} the discussion of strong decays of tetraquarks is revisited.
Mechanism of Fig. 1.b is mentioned in the end of the work but is not
systematically evaluated for all decay modes. This has been the motivation
of \cite{tq}, where the inclusions of both decay modes of Fig. 1.a and Fig.
1.b in a $SU_{V}(3)$-invariant interaction Lagrangian parameterized by two
coupling constants $c_{1}$ and $c_{2}$, is performed in all decay channels.

\bigskip

\begin{figure}[tbp]
\vspace*{-.5cm} \epsfig{figure=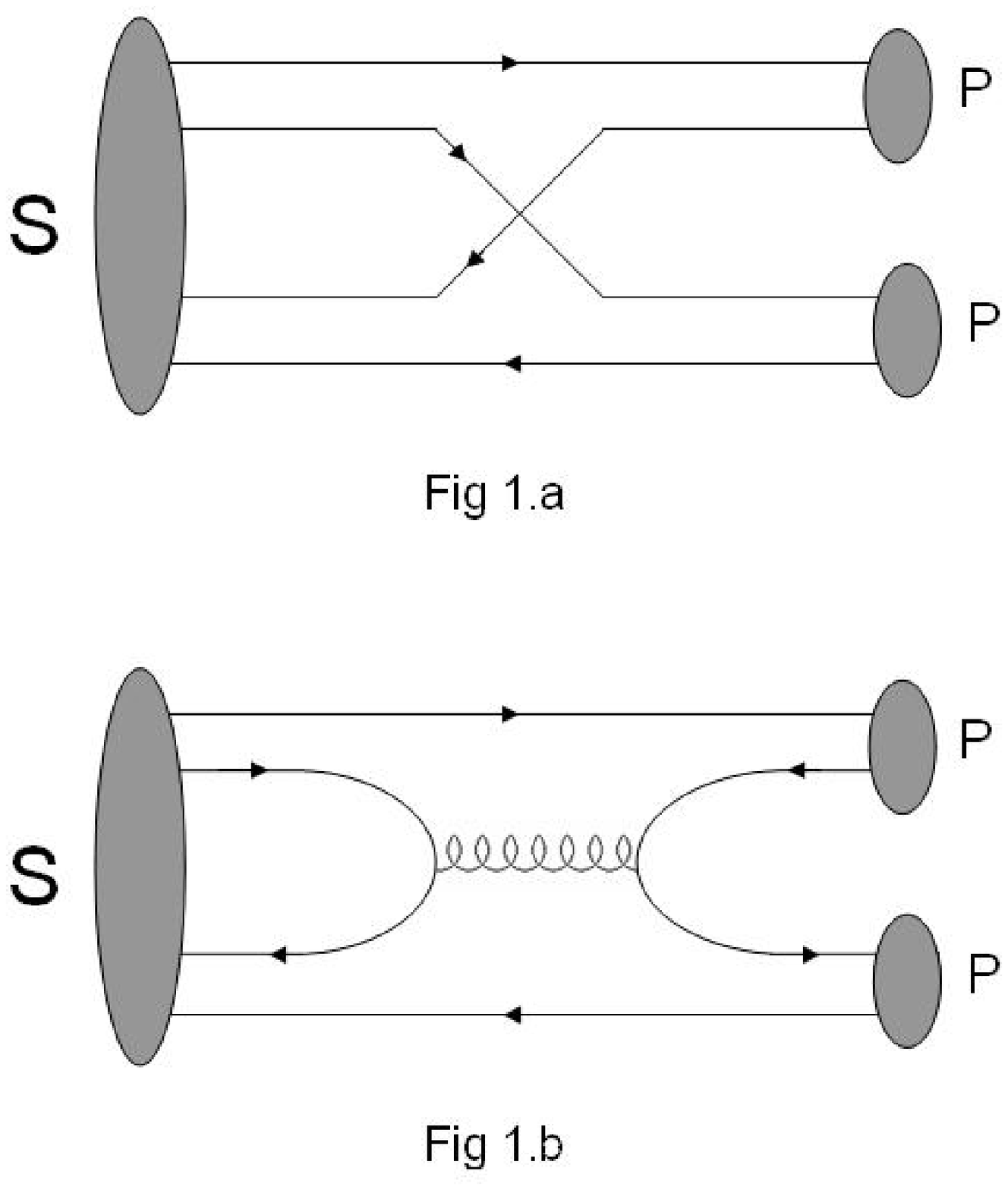,scale=.55} \vspace*{-.5cm} \centering%
{\ }
\caption{{Dominant (1.a) and subdominant (1.b) contributions to the
transition amplitudes of a scalar tetraquark state into two pseudoscalar
mesons.}}
\label{Fig1}
\end{figure}

The nonet of scalar tetraquark is described by the matrix \cite{tq}:%
\begin{eqnarray}
\mathcal{S}^{[4q]} &\equiv &\frac{1}{2}\left( 
\begin{array}{ccc}
\lbrack \overline{d},\overline{s}][d,s] & -[\overline{d},\overline{s}][u,s]
& [\overline{d},\overline{s}][u,d] \\ 
-[\overline{u},\overline{s}][d,s] & [\overline{u},\overline{s}][u,s] & -[%
\overline{u},\overline{s}][u,d] \\ 
\lbrack \overline{u},\overline{d}][d,s] & -[\overline{u},\overline{d}][u,s]
& [\overline{u},\overline{d}][u,d]%
\end{array}%
\right)  \label{s4q1} \\
&=&\left( 
\begin{array}{ccc}
\sqrt{\frac{1}{2}}(f_{B}-a_{0}^{0}(980)) & -a_{0}^{+}(980) & k^{+} \\ 
-a_{0}^{-}(980) & \sqrt{\frac{1}{2}}(f_{B}+a_{0}^{0}(980)) & -k^{0} \\ 
k^{-} & -\overline{k}^{0} & \sigma _{B}%
\end{array}%
\right)  \label{s4q2}
\end{eqnarray}%
where the notation $[u,d]$ refers to the scalar diquark with flavor
antysymmetric wave function $(ud-du).$ Thus, the tetraquark composition of
the light scalars is directly readable from Eqs. (\ref{s4q1})-(\ref{s4q2}).
A mixing of the isoscalar tetraquark states $\sigma _{B}[4q]\equiv \sigma
_{B}$ and $f_{B}[4q]\equiv f_{B}$, leading to the physical states $%
f_{0}(600) $ and $f_{0}(980)$, occurs \cite{tq}. The pseudoscalar nonet is
contained in the usual matrix $\mathcal{P}=\frac{1}{\sqrt{2}}%
\sum_{i=0}^{8}P^{i}\lambda _{i}$. The parity, charge conjugation and flavor
invariant interaction term reads 
\begin{equation}
\mathcal{L}_{\mathcal{S}^{[4q]}PP}=c_{1}\mathcal{S}_{ij}^{[4q]}Tr\left[ A^{j}%
\mathcal{P}^{t}A^{i}\mathcal{P}\right] -c_{2}\mathcal{S}_{ij}^{[4q]}Tr\left[
A^{j}A^{i}\mathcal{P}^{2}\right]  \label{lint}
\end{equation}%
with\ \ $\left( A^{i}\right) _{jk}=\varepsilon _{ijk}$. The first term,
proportional to $c_{1}$, describes the dominant, fall-apart
(OZI-superallowed mechanism) of Fig. 1.a and the second term, proportional
to $c_{2}$, the subdominant mechanism of Fig. 1.b.

The decay rates for the channel $S\rightarrow P_{1}P_{2}$ ($S$ stands for
scalar, $P$ for pseudoscalar) in the tree-level approximation reads 
\begin{equation}
\Gamma _{S\rightarrow P_{1}P_{2}}=\frac{p_{S\rightarrow P_{1}P_{2}}}{8\pi
M_{S}^{2}}g_{SP_{1}P_{2}}^{2},  \label{gsp1p2}
\end{equation}%
where $p_{S\rightarrow P_{1}P_{2}}$ is the three-momentum of (one of) the
outgoing particle(s). The coupling constants $g_{SP_{1}P_{2}}$ are function
of the decay constants $c_{1}$ and $c_{2}$ (symmetry factors included).
Formula (\ref{gsp1p2}) is strictly valid only for narrow states with $%
M_{S}>>M_{P_{1}}+M_{P_{2}};$ it serves, however, as a useful mnemonic for
the used conventions. The amplitude $g_{SP_{1}P_{2}}$ characterizes the
strength in a given channel and can be (in principle) extracted directly
from experiment. While full values are still unclear, ratios of coupling
constants are more stable. In particular, in \cite{bugg} the following ratio
is reported:%
\begin{equation}
\frac{g_{f_{0}(980)\overline{K}K}^{2}}{g_{a_{0}(980)\overline{K}K}^{2}}%
=2.15\pm 0.4.  \label{exp}
\end{equation}

The theoretical amplitudes in the $\overline{K}K$ channel for $f_{B},$ which
in the limit of zero mixing angle coincides with $f_{0}(980)$, and $%
a_{0}(980)$ as extracted from Eq. (\ref{lint}) read:%
\begin{eqnarray}
g_{f_{B}\overline{K}K} &=&\sqrt{2}\left( \sqrt{2}c_{1}+\frac{3}{\sqrt{2}}%
c_{2}\right) , \\
g_{a_{0}(980)\overline{K}K} &=&\sqrt{2}\left( \sqrt{2}c_{1}+\frac{1}{\sqrt{2}%
}c_{2}\right) .
\end{eqnarray}

If $c_{2}=0$ the ratio $g_{f_{B}\overline{K}K}^{2}/g_{a_{0}(980)\overline{K}%
K}^{2}$ is 1 and thus not in agreement with Eq. (\ref{exp}). When including
mixing in the isoscalar sector, which however should be small in order not
to spoil the mass degeneracy of $a_{0}(980)$ and $f_{0}(980)$, the situation
is worsened: the ratio turns out to be smaller than 1.

On the contrary, one can notice that the subdominant decay mechanism
enhances $g_{f_{B}\overline{K}K}$ more than $g_{a_{0}(980)\overline{K}K}.$
Even a relatively small $c_{2}$ can account for the value in Eq. (\ref{exp}%
). In \cite{tq} a fit to ratios of coupling constants allowed to fix (in a
possible mixing scenario) $c_{2}/c_{1}=0.62,$ smaller than $1$ but slightly
larger than $1/N_{c}=1/3.$ It is however important to say that within the
mentioned fit the coupling of $f_{0}(600)$ to $\overline{K}K$ turns out to
be small, while it is found to be sizable in \cite{bugg}. We refer to \cite%
{tq} for a more detailed analysis of admissible solutions, but more work is
needed. For instance, fits to the line shapes of $\phi $ decays (even using
different parametrizations than the kaon loop model, see for instance Ref. 
\cite{achasov} and Refs. therein) should be performed with new data from
KLOE experiment \cite{kloe}. If the ratio (\ref{exp}) shall be confirmed,
the inclusion of the subdominant decay mode is necessary for tetraquark
picture of light scalars.

\section{Going further: mixing and loops}

\subsection{Mixing of tetraquark and quarkonia states in a chiral approach}

Assuming a nonet of tetraquark states below 1 GeV, mixing with a nonet of
quarkonia (with glueball) above 1 GeV takes place and the physical states
turn out to be an admixture of a tetraquark and a quarkonium. Thus, in this
sense the identification of Eq. (\ref{s4q1}) with Eq. (\ref{s4q2}) is not
valid any longer.

A crucial point is to estimate the intensity of tetraquark-quarkonia mixing 
\cite{tqmix}. If it is large, the study of decays and branching ratios is
largely influenced and the attempt to explain some phenomenological
features, such as the ratio of coupling constants, with only one dominant
assignment cannot be correct and both components shall be considered. On the
contrary, if it is small it is possible to perform, for instance, the study
of the previous section.

In order to study the mixing of tetraquark and quarkonia we have to enlarge
the symmetry group. Instead of only $SU_{V}(3)$ invariance as in Eq. (\ref%
{lint}), full chiral symmetry $SU_{V}(3)\times SU_{A}(3)$ has to be
considered. Let us introduce, together with the pseudoscalar quarkonia nonet
matrix $\mathcal{P},$ the scalar quarkonia nonet matrix $\mathcal{S}^{[%
\overline{q}q]}$. We will concentrate on non-isosinglet channels, thus we
can disregard the glueball in the following. As usual the complex matrix $%
\Sigma =\mathcal{S}^{[\overline{q}q]}+i\mathcal{P}$ is introduced. The
following chiral Lagrangian

\begin{equation}
\mathcal{L}_{\Sigma }=\frac{1}{4}Tr\left[ \partial _{\mu }\Sigma \partial
^{\mu }\Sigma ^{\dag }\right] -V_{0}(\Sigma ,\Sigma ^{\dag })-V_{SB}(\Sigma
,\Sigma ^{\dag })  \label{lsigma}
\end{equation}%
($Tr$ denotes trace over flavor) describes the dynamics of the pseudoscalar
and scalar quarkonia mesons, where $V_{0}$ represents the chiral invariant
potential while $V_{SB}$ encodes symmetry breaking due to the non-zero
current quark masses. It is important to stress that we do $not$ need to
know the potentials $V_{0}$ and $V_{SB}$. What is important for us is simply
that spontaneous chiral symmetry breaking ($\chi SB$) occurs. The minimum $%
\Sigma _{0}$ of the potential $V_{0}+V_{SB}$ is non-zero and is given in a
model independent way by the following vacuum expectation values ($vev$'s) 
\cite{shechter,geffen}:%
\begin{equation}
\Sigma _{0}=diag\left\{ \frac{F_{\pi }}{\sqrt{2}},\frac{F_{\pi }}{\sqrt{2}},%
\sqrt{2}F_{K}-\frac{F_{\pi }}{\sqrt{2}}\right\} =diag\left\{ \alpha
_{u},\alpha _{u},\alpha _{s}\right\}   \label{csb1}
\end{equation}%
where $F_{\pi }$ and $F_{K}$ are the pion and kaon decay constants.

The full chirally invariant interaction Lagrangian involves not only scalar,
but also pseudoscalar diquarks \cite{tqmix}. The latter are however more
massive and less bound \cite{psdiq}. Retaining only the scalar `good'
diquarks, the interaction Lagrangian, connecting the tetraquark nonet $%
\mathcal{S}^{[4q]}$ to the quarkonia states $\Sigma =\mathcal{S}^{[\overline{%
q}q]}+i\mathcal{P}$ and taking into account diagrams of the type of Fig. 1.a
and 1.b, turns out to be \cite{tqmix}:

\begin{equation}
\mathcal{L}_{\mathcal{S}^{[4q]}\Sigma \Sigma }=-\frac{c_{1}}{2}\mathcal{S}%
_{ij}^{[4q]}Tr\left[ A^{j}\Sigma ^{t}A^{i}\Sigma +D\Sigma ^{\ast }D^{\dagger
}\Sigma ^{\dagger }\right] -\frac{c_{2}}{2}\mathcal{S}_{ij}^{[4q]}Tr\left[
A^{j}A^{i}\Sigma ^{\dagger }\Sigma +A^{j}A^{i}\Sigma \Sigma ^{\dagger }%
\right] .  \label{lint2}
\end{equation}%
Notice that the coupling constants $c_{1}$ and $c_{2}$ are the same of the
previous section: in fact, $\mathcal{L}_{\mathcal{S}^{[4q]}PP}$ of Eq. (\ref%
{lint}) is entirely contained $\mathcal{L}_{\mathcal{S}^{[4q]}\Sigma \Sigma
}.$ When expanding around the minimum $\Sigma _{0}$ of Eq. (\ref{csb1})
mixing among scalar tetraquark and quarkonia arises: the mixing strength in
a given channel is then a combination of the two couplings $c_{1},$ $c_{2}$
and the decay constants $F_{\pi },$ $F_{K}.$ Thus, using experimental
informations on decay widths of scalar states from PDG \cite{pdg}, which
constraints $c_{1}$ and $c_{2},$ we can estimate the intensity of mixing.
Let us make a concrete example in the neutral isovector channel $I=1$. The
mixing term emerging from the interaction Lagrangian (\ref{lint2}) and from
equation (\ref{csb1}) reads%
\begin{equation}
\mathcal{L}_{\mathcal{S}^{[4q]}\Sigma \Sigma }\rightarrow \mathcal{L}%
_{mixa_{0}^{0}}=2(c_{1}\alpha _{s}+c_{2}\alpha _{u})\left(
a_{0}^{0}[4q]\cdot a_{0}^{0}[\overline{q}q]\right) 
\end{equation}%
where $a_{0}^{0}[4q]=\frac{1}{2}([\overline{u},\overline{s}][u,s]-[\overline{%
d},\overline{s}][d,s])$ refers to a bare tetraquark state and $a_{0}^{0}[%
\overline{q}q]=\frac{1}{\sqrt{2}}(\overline{u}u-\overline{d}d)$ to the
quarkonium. The physical resonances $a_{0}(980)$ and $a_{0}(1450)$ are then
an admixture of the bare components:%
\begin{equation}
\left( 
\begin{array}{c}
a_{0}^{0}(980) \\ 
a_{0}^{0}(1470)%
\end{array}%
\right) =\left( 
\begin{array}{cc}
\cos (\theta ) & \sin (\theta ) \\ 
-\sin (\theta ) & \cos (\theta )%
\end{array}%
\right) \left( 
\begin{array}{c}
a_{0}^{0}[4q] \\ 
a_{0}^{0}[\overline{q}q]%
\end{array}%
\right) .
\end{equation}%
Bounds on the full decay widths from PDG allow us to reach the conclusion $%
(\sin \theta )^{2}<10\%,$ that is the mixing angle is small, see details in 
\cite{tqmix}. The state $a_{0}^{0}(980)\simeq a_{0}^{0}[4q]$ is dominantly a
tetraquark and $a_{0}^{0}(1470)\simeq a_{0}^{0}[\overline{q}q]$ a
quarkonium. A similar result holds for the kaonic states, see however also
the different outcome in \cite{fariborz}. In turn, this means that
identifying Eq. (\ref{s4q1}) with Eq. (\ref{s4q2}), as done in the previous
section, is correct in first approximation. Notice moreover that these
results depend only very weakly on the ratio $c_{2}/c_{1}.$

A note of care concerning the isoscalar sector, where a direct mixing of 5
states takes place (two tetraquarks two quarkonia and one glueball), is
needed . While results are consistent with a substantial separation of
tetraquark form quarkonia (plus glueball) states, my point of view is that
experimental information is not yet good enough to unequivocally extract the
mixing angles.

As a last step we mention the results concerning the emerging of a
tetraquark condensate: the field $\sigma _{B}[4q]$, referring to $\frac{1}{2}%
[\overline{u},\overline{d}][u,d]$ and introduced in Eqs. (\ref{s4q1}) and (%
\ref{s4q2}), acquires a nonzero $vev$ \cite{tqmix}:%
\begin{equation}
\left\langle \sigma _{B}[4q]\right\rangle =\frac{c_{1}+c_{2}}{M_{\sigma
_{B}[4q]}^{2}}F_{\pi }^{2}
\end{equation}%
where $M_{\sigma _{B}[4q]}\simeq 0.6$ GeV. Notice that this $vev$ vanishes
if $F_{\pi }\rightarrow 0,$ i.e. if the quark condensate is zero: the
tetraquark condensate is induced by the quark condensate in this framework.
The actual value of the tetraquark condensate can be estimated by
introducing $\Lambda _{QCD}\sim 0.25$ GeV as in \cite{shechter} $%
\left\langle \frac{1}{2}[\overline{u},\overline{d}][u,d]\right\rangle \sim
\Lambda _{QCD}^{5}\left\langle \sigma _{B}[4q]\right\rangle \simeq 3.1\cdot
10^{-5}$ GeV$^{6}$.

\subsection{Mesonic loops}

All the considerations up to now were based on the tree-level approximation.
The inclusion of mesonic loops and finite-width effects is the next step in
the study of decays of light scalars \cite{lupo}. Be the propagator of a
scalar state $S$ the function $\Delta _{S}(x)=\left( x^{2}-M_{0}^{2}+\Pi
(x)+i\varepsilon \right) ^{-1}$ where $M_{0}$ is the bare mass, $\Pi (x)$ is
the dressing function of mesonic loops (for instance, $\pi \pi $ and $%
\overline{K}K$ loops contribute for $f_{0}(980)$) and $x=\sqrt{p^{2}}.$ The
spectral function is defined as $d_{S}(x)=\frac{2x}{\pi }\lim_{\varepsilon
\rightarrow 0}\Delta _{S}(x).$ When the propagator satisfy the K\"{a}%
llen-Lehman representation the normalization $\int_{0}^{\infty }d_{S}(x)%
\mathrm{dx}=1$ is assured \cite{achasovprop}. The function $d_{S}(x)$ is the
mass-distribution of the scalar state $S.$ The real part of the loop
function $\Pi (x)$ is made convergent by `delocalizing' the corresponding
interaction Lagrangian by a cutoff function which dies off at a scale of
1.5-2 GeV. All quantities are evaluated in the scheme of a nonlocal
Lagrangian and the normalization of $d_{S}(x)$ is a consequence (and a
consistency check) of the calculation, see details in \cite{lupo}. The decay
formula (\ref{gsp1p2}) modifies as follows:%
\begin{equation}
\Gamma _{S\rightarrow P_{1}P_{2}}=\int_{0}^{\infty }\mathrm{dx}\frac{%
p_{S\rightarrow P_{1}P_{2}}}{8\pi x^{2}}g_{SP_{1}P_{2}}^{2}d_{S}(x)
\label{gfull}
\end{equation}%
where the average over the mass-distribution $d_{S}(x)$ is evident. Some
comments are in order: (a) the mass of the scalar particle can be defined as
the zero of the real part of the propagator, $M^{2}-M_{0}^{2}+$Re$[\Pi
(M)]=0 $. Typically $M<M_{0},$ i.e. quantum corrections lower the mass.
Various definitions of mass (including the widely used position of the pole
on the complex plane) are possible. Indeed, the average-mass $\left\langle
M\right\rangle =$ $\int_{0}^{\infty }\mathrm{dx}d_{S}(x)x$ can be an
intuitive alternative. (b) The width, as evaluated via Eq. (\ref{gfull}), is
(slightly) smaller than the tree-level counterpart far from threshold. For
the $\overline{K}K$ mode of $a_{0}(980)$ and $f_{0}(980)$ it is, on the
contrary, sizable while the tree-level decay rate vanishes. (c) The states $%
a_{0}(980)$ and $f_{0}(980)$ turn out to contain large amount of $\overline{K%
}K$ clouds, i.e. they spend consistent part of their life as $\overline{K}K$
pairs. On a microscopic level the scalar diquarks are extended objects with
a radius of about 0.2-0.5 fm (a small radius is found within the NJL model 
\cite{suzuki} while a larger one within the BS-approach \cite{roberts}).
Being the typical meson radius of about 0.5-1 fm it is clear that a
non-negligible overlap of the two diquarks takes place. As depicted in Fig.
1.a the rearrangement of the two colored bumps (diquark) into two colorless
mesons occurs: the system spends consistent part of its life as a mesonic
molecular state \cite{hanhart}.

Although all these effects are important and deserve further study such as
the inclusion of derivative coupling and $a_{0}(980)$-$f_{0}(980)$ mixing
effects \cite{hanhartmix}, the scenario as outlined in Section 1 is not
qualitatively modified: Clebsch-Gordan coefficients of tree-level amplitudes
offer a way to distinguish among different spectroscopic interpretations of
the light scalars.

\section{Conclusions}

In this paper we studied some aspects of the tetraquark interpretation of
light scalars: the next to leading order in the large-$N_{c}$ expansion, see
Fig. 1.b, has been introduced and its consequences briefly summarized in
Section 2 \cite{tq}. The decisive point in favour of Fig. 1.b is the ratio 
\cite{bugg} $g_{f_{0}(980)\overline{K}K}^{2}/g_{a_{0}(980)\overline{K}%
K}^{2}=2.15\pm 0.4$. Future studies are needed to extract precisely the
values of the coupling constants, thus allowing to compare different
assignments for the light scalars.

We then turned the attention to two related phenomena: (a) the mixing of
tetraquark and quarkonia states, studied in the framework of a chiral
Lagrangian \cite{tqmix}: the results point to a small mixing and to a
separation of tetraquark states ($<$ 1 GeV) and quarkonia ($>$ 1 GeV).
Noticeably, one does not need to specify the form of the chiral Lagrangian:
the requirement of chiral symmetry breaking, together with the experimental
information on decay widths from \cite{pdg}, is sufficient to find upper
limits for the mixing angles. (b) Mesonic loops and finite-width effects
represent an important extension beyond tree-level \cite{lupo} allowing the
evaluation $a_{0}(980)\rightarrow \overline{K}K$ and $f_{0}(980)\rightarrow 
\overline{K}K$. Both extensions (a) and (b) do not change the qualitative
picture emerging in Section 2.

As mentioned in the Introduction in the recent unquenched calculation of 
\cite{graznew} the $a_{0}(1450)$ is still present, but a lower state,
compatible with $a_{0}(980)$ shows up: however, unquenched lattice
calculations $do$ include quark loops and therefore induce also tetraquark
admixtures. Indeed, in the ideal limit Lattice simulations should reproduce
the physical states regardless of their inner structure. In this sense
quenched simulations might be more suitable to answer the question related
to the bare quarkonia states: for this reason we still consider valid the
results of \cite{quenched}, according to which the bare isovector quarkonium
lies in the $1.5$ GeV mass region.

The two-photon mechanism has not been discussed here but will be subject of
a forthcoming proceeding, see however \cite{tq} and the recent work \cite%
{sgg}. Future theoretical studies aiming to understand the nature of light
scalars will focus on the following subjects:\ (i) the radiative decay of $%
\phi $ meson ($\phi \rightarrow \gamma S\rightarrow \gamma PP)$, in
particular concerning the line shapes \cite{kloe}. (ii) Predictions for
scalar decays into a vector meson and a photon ($S\rightarrow V\gamma )$ 
\cite{kala}. (iii) Decays of heavier states, such as $J/\psi $, into light
scalars taking into account the tetraquark interpretation of the latter.

\end{document}